\documentclass{article}%
\usepackage{amssymb}
\usepackage{amsmath}
\usepackage{amsfonts}
\usepackage{graphicx}%
\providecommand{\U}[1]{\protect\rule{.1in}{.1in}}

\begin{document}
		\let\vaccent=\v 
		\renewcommand{\v}[1]{\ensuremath{\mathbf{#1}}} 
		\newcommand{\gv}[1]{\ensuremath{\mbox{\boldmath$ #1 $}}}    
		\newcommand{\uv}[1]{\ensuremath{\mathbf{\hat{#1}}}} 
		\newcommand{\abs}[1]{\left| #1 \right|} 
	   \newcommand{\norm}[1]{\left|\left| #1 \right|\right|} 
	\newcommand{\inner}[2]{\left<#1 \vphantom{#2}\right, \left.#2\vphantom{#1}\right>} 
		\newcommand{\avg}[1]{\left< #1 \right>} 
		\let\underdot=\d 
		\renewcommand{\d}[2]{\frac{d #1}{d #2}} 
		\newcommand{\dd}[2]{\frac{d^2 #1}{d #2^2}} 
		\newcommand{\pd}[2]{\frac{\partial #1}{\partial #2}} 
		\newcommand{\pdd}[2]{\frac{\partial^2 #1}{\partial #2^2}} 
		\newcommand{\pddd}[2]{\frac{\partial^3 #1}{\partial #2^3}} 
		\newcommand{\pdc}[3]{\left( \frac{\partial #1}{\partial #2}
		 \right)_{#3}} 
		\newcommand{\ket}[1]{\left| #1 \right>} 
		\newcommand{\bra}[1]{\left< #1 \right|} 
		\newcommand{\braket}[2]{\left<#1\vphantom{#2}\right| \left.#2\vphantom{#1}\right>} 
		\newcommand{\ketbra}[2]{\left|#1\vphantom{#2}\right> \left<#2\vphantom{#1}\right|} 
		\newcommand{\matrixel}[3]{\left< #1 \vphantom{#2#3} \right| #2 \left| #3 \vphantom{#1#2} \right>} 
		\newcommand{\commute}[2]{\left[#1,#2\right]} 
         \newcommand{\anticommute}[2]{\left\{#1,#2\right\}} 
		\newcommand{\trace}[1]{\mathrm{Tr}\left[#1\right]} 
\newcommand{\parttrace}[2]{\mathrm{Tr}_{#1}\left[#2\right]} 
		\newcommand{\grad}[1]{\gv{\nabla} #1} 
		\let\divsymb=\div 
		\renewcommand{\div}[1]{\gv{\nabla} \cdot #1} 
		\newcommand{\curl}[1]{\gv{\nabla} \times #1} 
		\newcommand{\properint}[4]{\int_{#1}^{#2}#3\mathrm{d}#4}
		\newcommand{\improperint}[2]{\int_{-\infty}^{+\infty}#1\mathrm{d}#2}
		\newcommand{\intB}[3]{\lim_{x^-\rightarrow 0}\int \mathrm{d}#1 \ \mathrm{d}#2  \  #3}
		\newcommand{\intBs}[3]{\lim_{s\rightarrow 0}\int \mathrm{d}#1 \ \mathrm{d}#2  \  #3}
		\newcommand{\intQ}[3]{\int \mathrm{d}#1 \ \mathrm{d}#2  \  #3}
		\newcommand*\Laplace{\mathop{}\!\mathbin\bigtriangleup}
\newcommand*\DAlambert{\mathop{}\!\mathbin\Box}
		
	

\newcommand{\re}[1]{\mathcal{R}\mathrm{e} \left[#1\right] }
\newcommand{\im}[1]{\mathcal{I}\mathrm{m} \left[#1\right] }

\begin{titlepage}
\ \\
\begin{center}
\LARGE
{\bf
Pure State Entanglement  Harvesting \\ in Quantum Field Theory 
}
\end{center}
\ \\
\begin{center}
\large{Jose Trevison, Koji Yamaguchi and Masahiro Hotta }\\
{\it
Graduate School of Science, Tohoku University,\\ Sendai, 980-8578, Japan}
\end{center}
\begin{abstract}
Quantum fields in vacuum states carry an infinite amount of quantum entanglement, and its entanglement entropy has the ultraviolet divergence. Harvesting protocols of the vacuum entanglement had been investigated, but its efficiency is very low to date. The main reason of the low efficiency originates from the fact that the extracted entanglement is embedded in a mixed state of two external devices. We propose a general protocol with high efficiency by extracting pure state entanglement from the field. We only use bi-linear interactions between the fields and external devices. Even though the ultraviolet cutoff remains finite, the protocol is capable of extracting a huge amount of entanglement. Hence the infinite amount of entanglement extraction is attained without the ultraviolet divergence of the field entanglement entropy in a continuum limit. There exists  a trade-off relation between the extracted entanglement and its energy cost.
\end{abstract}
\end{titlepage}

\ 

\section{\bigskip Introduction}

Quantum entanglement is a key resource in a wide class of quantum protocols
including quantum computation. For two quantum systems $A$ and $B$ in a pure
state $|\psi\rangle_{AB}$, it is quantified by entanglement entropy, which is
defined by $S_{EE}=-\operatorname*{Tr}_{A}\left[  \rho_{A}\ln\rho_{A}\right]
$ with the reduced operator $\rho_{A}=\operatorname*{Tr}_{B}\left[
|\psi\rangle_{AB}\langle\psi|_{AB}\right]  $ of $A$. In quantum field theory,
$S_{EE}$ includes ultraviolet divergence \cite{EE1}-\cite{EE3}. In $D$
dimensional space-time with $D\geq3$, $S_{EE}$ between two adjacent regions
obeys the area law such that $S_{EE}\propto\left\vert \partial A\right\vert
/\epsilon^{D-2}$, where $\left\vert \partial A\right\vert $ is the boundary
area of $A$ and $\epsilon$ is ultraviolet cutoff. In $D=2$, $S_{EE}%
\varpropto\ln\left(  \left\vert A\right\vert /\epsilon\right)  $ holds, where
$\left\vert A\right\vert $ is width of $A$. By taking $\epsilon\rightarrow0$,
$S_{EE}$ tends to infinity. This fact makes entanglement harvesting protocols fascinating. 
In these protocols two external quantum systems coupled to a quantum
field in the vacuum state extract entanglement from the field \cite{EH1} \cite{EH2}. 
This considerable possibility is expected to be experimentally explored in various
systems including quantum optics with strong coupling to superconducting
circuits \cite{QO}\cite{SC}. The huge entanglement may enhance information
capacity when quantum information is imprinted to the field. Quantum Hall
edge currents may be promising to assess this impact because the system is
described by a chiral mass-less field theory \cite{QHS} and simultaneously has
high compatibility with semiconductor technology. A disadvantage of
entanglement harvesting protocols proposed so far \cite{EH1}-\cite{EH4} is low
entanglement gain. The main reason for this flaw is that the external systems
are not in a pure state but in a mixed state after the harvesting, and remain entangled with the
field. One way to avoid the flaw is to transfer large entanglement between two
field subsystems \textit{in a pure state} to the two external systems. The
pure state is a perfect state. The system in a pure state does not have any
correlation with other external systems, and the whole information about
physical quantities of the system is imprinted in the pure state. From this
point of view, the concept of purification partner in black hole physics
\cite{HSU} is promising in condensed matter physics. In \cite{HSU}, for a
Hawking particle emitted by a black hole, its partner particle was clearly
defined to be in a pure Gaussian state of the two particle system. This
partner definition provides a deep insight for the black hole information loss
problem \cite{h}. In this paper, we apply a generalized partner for an
arbitrarily local mode in an arbitrary states of a scalar field, which was
proposed in \cite{TYH}. There exist two distinct classes of the general
partners for local modes. One of them is ordinary, and called
\textit{spatially separated partner} (SSP). The local mode functions of SSP of
a field have no spatial overlap. Another is called \textit{spatially}
\textit{overlapped partner} (SOP). The local mode functions of SOP actually
have a spatial overlap. The two independent modes $A$ and $B$ of SOP are
defined by using an operator algebra which satisfies locality condition of
operators $O_{A}$ of $A$ and $O_{B}$ of $B$ such that $\left[  O_{A}%
,O_{B}\right]  =0$. This locality allows us to introduce quantum entanglement
between the spatially overlapped systems in the same philosophy of the
correlation space in quantum computation \cite{cc1} \cite{cc2}. Extraction
of SOP from a field in the vacuum state is able to yield pure-state
entanglement via simple bi-linear couplings between the field and outside
devices. We propose an entanglement harvesting protocol, which is capable of
extraction of unlimited amount of entanglement. It is worth stressing that
this happens even when the ultraviolet cutoff and the degrees of freedom
remain finite. Our pure state extraction permits entanglement generation during the process. In fact, merely three coupled harmonic oscillators as a
discretized field model with finite lattice spacing can afford to provide
unbounded amount of entanglement from the zero point fluctuation. Ultraviolet
divergence of the continuum field entanglement is not actually requested for
the huge entanglement harvesting via the bi-linear coupling. There exists a
trade-off relation between the extracted entanglement and its energy cost.
Infinite amount of entanglement extraction out of a field in the vacuum state
requires infinite energy to operate the devices for the harvesting.

In section 2, we review a general partner formula for a discretized free
scalar field in 1+1 dimensional space-time. In section 3, we introduce an
entanglement harvesting protocol for SOP extraction from the field. In section
4, we analyze energy cost of huge entanglement extraction. In section 5,
summary and discussion are provided.

In this paper, the natural unit is adopted: $c=\hbar=1$.

\bigskip

\section{Spatially Overlapped Partner}

In this section, we briefly review a general partner formula in 1+1
dimensional space-time. Let us suppose a free scalar quantum field
$\hat{\varphi}(x,t)$ with mass $m$. The space satisfies a periodic boundary
condition as $\hat{\varphi}(x+L,t)=\hat{\varphi}(x,t)$, where $L$ is the
entire space length. The Hamiltonian reads%

\begin{equation}
\hat{H}_{\varphi}=\frac{1}{2}\int_{-L/2}^{L/2}:\hat{\Pi}(x)^{2}:dx+\frac{1}{2}%
\int_{-L/2}^{L/2}:\left(  \partial_{x}\hat{\varphi}(x)\right)  ^{2}%
:dx+\frac{m^{2}}{2}\int_{-L/2}^{L/2}:\hat{\varphi}(x)^{2}:dx,
\end{equation}
where $\hat{\Pi}(x)$ is canonical momentum operator satisfying
\begin{equation}
\left[  \hat{\varphi}(x),\hat{\Pi}(x^{\prime})\right]  =i\delta\left(
x-x^{\prime}\right)  .
\end{equation}
The discretized model with lattice spacing $\epsilon$ is a system of coupled
harmonic oscillators. Each field operator corresponds to $\left(  \hat{q}%
_{n},\hat{p}_{n}\right)  $ satisfying $\left[  \hat{q}_{n},\hat{p}_{n^{\prime
}}\right]  =i\delta_{nn^{\prime}}$ in the following way: \bigskip%
\begin{align}
\hat{\varphi}(x)  &  \rightarrow\frac{\hat{q}_{n}}{\sqrt{m\epsilon}},\\
\hat{\Pi}(x)  &  \rightarrow\sqrt{\frac{m}{\epsilon}}\hat{p}_{n} \ .
\end{align}
By introducing dimensionless time variable $\tau=mt$, the Hamiltonian is given by
\begin{equation}
\hat{H}=\frac{1}{2}\sum_{n=1}^{N}:\hat{p}_{n}^{2}:+\left(  \frac{1}{2}+\eta\right)
\sum_{n=1}^{N}:\hat{q}_{n}^{2}:-\eta\sum_{n=1}^{N}:\hat{q}_{n+1}\hat{q}_{n}:,
\label{H}
\end{equation}
where $N=L/\epsilon$ and $\eta=\left(  m\epsilon\right)  ^{-2}$. Using orthonormal mode functions defined as%
\begin{equation}
u_{k}(n)=\frac{1}{\sqrt{N}}\exp\left[  i\left(  2\pi k\frac{n}{N}\right)
\right]  ,
\end{equation}
$\hat{q}_{n}$ and $\hat{p}_{n}$ are expanded as%
\begin{equation}
\hat{q}_{n}=\sum_{k=0}^{N-1}\frac{1}{\sqrt{2\omega_{k}}}\left[  \hat{a}_{k}%
u_{k}(n)+\hat{a}_{k}^{\dag}u_{k}(n)^{\ast}\right]  ,
\end{equation}
\begin{equation}
\hat{p}_{n}=\frac{1}{i}\sum_{k=0}^{N-1}\sqrt{\frac{\omega_{k}}{2}}\left[
\hat{a}_{k}u_{k}(n)-\hat{a}_{k}^{\dag}u_{k}(n)^{\ast}\right]  ,
\end{equation}
where $\left[  \hat{a}_{k},\hat{a}_{k^{\prime}}^{\dag}\right]  =\delta_{kk^{\prime}}$. The
dimensionless discretized energy for momentum $k$ is given by%
\begin{equation}
\omega_{k}^{2}=1+2\eta\left[  1-\cos\left(  \frac{2\pi k}{N}\right)  \right] \ . 
\end{equation}
The vacuum state $|0\rangle$ is defined as $\hat{a}_{k}|0\rangle=0$. The state is a Gaussian state which is governed by the two following correlation functions:%
\begin{align}
\Delta_{q}(n-n^{\prime})\equiv \langle0|\hat{q}_{n}\hat{q}_{n^{\prime}}|0\rangle &  =\frac{1}{N}\sum
_{k=0}^{N-1}\frac{1}{2\omega_{k}}\cos\left(  2\pi\frac{k\left(  n-n^{\prime
}\right)  }{N}\right)  ,
\label{corrq}
\\
\Delta_{p}(n-n^{\prime})\equiv \langle0|\hat{p}_{n}\hat{p}_{n^{\prime}}|0\rangle &  =\frac{1}{N}\sum
_{k=0}^{N-1}\frac{\omega_{k}}{2}\cos\left(  2\pi\frac{k\left(  n-n^{\prime
}\right)  }{N}\right)  .
\label{corrp}
\end{align}
The wave-function of the vacuum state is given by%
\begin{equation}
\Psi\left(  q_{1},\cdots,q_{N}\right)  =\langle q_{1},\cdots,q_{N}%
|0\rangle\varpropto\exp\left[  -\sum_{n=1}^{N}\sum_{n^{\prime}=1}^{N}%
q_{n}\Delta_{p}(n-n^{\prime})q_{n^{\prime}}\right]  .
\end{equation}
Let us suppose a local mode $A$ of the field defined by a canonical pair $\left(  \hat{q}_A,\hat{p}_A\right)  $ satisfying $\left[\hat{q}_A,\hat{p}_A\right]  =i$. The canonical pair is given by
\begin{align}
\hat{q}_A  &  =\sum_{n=1}^{N}\left(  x_A(n)\hat{q}_{n}+y_A(n) \hat{p}_{n}\right)  ,
\label{qA}
\\
\hat{p}_A  &  =\sum_{n=1}^{N}\left(  z_A(n)\hat{q}_{n}+w_A(n)\hat{p}_{n}\right)  
\label{pA} ,
\end{align}
where $\left(  x_A(n),y_A(n),z_A(n),w_A(n)\right)  $ are arbitrarily fixed real coefficients obeying
\begin{equation}
\sum_{n=1}^{N}\left(  x_A(n)w_A(n)-z_A(n)%
y_A(n)\right)  =1.
\label{comm}
\end{equation}
Applying a symplectic group transformation as
\begin{equation}
\left[
\begin{array}
[c]{c}%
\hat{Q}_{A}\\
\hat{P}_{A}%
\end{array}
\right]  =\left[
\begin{array}
[c]{cc}%
\cos\theta_{A} & \sin\theta_{A}\\
-\sin\theta_{A} & \cos\theta_{A}%
\end{array}
\right]  \left[
\begin{array}
[c]{cc}%
e^{\sigma_{A}} & 0\\
0 & e^{-\sigma_{A}}%
\end{array}
\right]  \left[
\begin{array}
[c]{cc}%
\cos\theta_{A}^{\prime} & \sin\theta_{A}^{\prime}\\
-\sin\theta_{A}^{\prime} & \cos\theta_{A}^{\prime}%
\end{array}
\right]  \left[
\begin{array}
[c]{c}%
\hat{q}_A\\
\hat{p}_A%
\end{array}
\right]  ,
\end{equation}
with appropriately fixed real parameters $\theta_{A},\theta_{A}^{\prime},\sigma_{A}$, we get the standard canonical pair $\left(  \hat{Q}_{A},\hat{P}_{A}\right)  $ of the mode, which yields the standard form of covariance matrix as
\begin{eqnarray}
M_{A}&=&\left[
\begin{array}
[c]{cc}%
\langle0|\hat{Q}_{A}^{2}|0\rangle & \frac{1}{2}\langle0|\hat{P}_{A}\hat{Q}%
_{A}+\hat{Q}_{A}\hat{P}_{A}|0\rangle\\
\frac{1}{2}\langle0|\hat{P}_{A}\hat{Q}_{A}+\hat{Q}_{A}\hat{P}_{A}|0\rangle &
\langle0|\hat{P}_{A}^{2}|0\rangle
\end{array}
\right] 
\nonumber
\\
 &=&\left[
\begin{array}
[c]{cc}%
\frac{1}{2}\sqrt{1+g^{2}} & 0\\
0 & \frac{1}{2}\sqrt{1+g^{2}}%
\end{array}
\right]  ,
\end{eqnarray}
where $g$ is a non-negative parameter and $\langle0|\hat{Q}_{A}\hat{P}_{A}|0\rangle=\frac{i}{2}$ holds. Let us expand the standard canonical pair in terms of $\hat{a}_{k}$ and $\hat{a}_{k}^{\dag}$ as
\begin{equation}
\hat{Q}_{A}=\left(  \frac{\sqrt{1+g^{2}}}{2}\right)  ^{1/2}\sum_{k=0}%
^{N-1}\left(  Q_A(k)^{\ast}\hat{a}_{k}+Q_A(k){}\hat{a}_{k}^{\dag}\right)  ,
\end{equation}
\begin{equation}
\hat{P}_{A}=\left(  \frac{\sqrt{1+g^{2}}}{2}\right)  ^{1/2}\sum_{k=0}%
^{N-1}\left(  P_{A}(k)^{\ast}\hat{a}_{k}+P_{A}(k){}\hat{a}_{k}^{\dag}\right)  .
\end{equation}
The coefficients $Q_A(k)$ and $P_{A}(k)$ satisfy the following conditions:
\begin{eqnarray}
\sum_{k=0}^{N-1}Q_A(k)^{\ast}Q_A(k){}&=&1,
\\
\sum_{k=0}^{N-1}P_{A}(k)^{\ast}P_{A}(k){}&=&1,
\\
\sum_{k=0}^{N-1}P_{A}(k)^{\ast}Q_A(k){}&=&-\frac{i}{\sqrt{1+g^{2}}}.
\end{eqnarray}

Let us define the partner mode $B$ of the mode $A$ by another canonical pair $\left(  \hat{Q}_{B},\hat{P}_{B}\right)  $:
\begin{align}
\hat{Q}_{B} &  = \left(  \frac{\sqrt{1+g^{2}}}{2}\right)  ^{1/2}\sum_{n=1}^{N}\left( X_B(n)\hat{q}_{n}+ Y_B(n) \hat{p}_{n}\right)  ,\\
\hat{P}_{B} &  = \left(  \frac{\sqrt{1+g^{2}}}{2}\right)  ^{1/2}\sum_{n=1}^{N}\left(  Z_B(n)\hat{q}_{n}+W_B(n)\hat{p}_{n}\right)  ,
\end{align}
which satisfies $\left[  \hat{Q}_{B},\hat{P}_{B}\right]  =i$. By measuring the covariance matrix of $A$ and $B$  given by
\begin{align}
&  M_{AB}
 =\left(
\begin{array}
[c]{cccc}%
\langle0|\hat{Q}_{A}^{2}|0\rangle & 
\re{\matrixel{0}{\hat{Q}_A\hat{P}_A}{0}}  &
 \langle0|\hat{Q}_{A}\hat{Q}_{B}|0\rangle &
  \langle0|\hat{Q}_{A}\hat{P}_{B}|0\rangle
  \\
\re{\matrixel{0}{\hat{Q}_A\hat{P}_A}{0}} &
\langle0|\hat{P}_{A}^{2}|0\rangle &
 \langle0|\hat{P}_{A}\hat{Q}_{B}|0\rangle &
\langle0|\hat{P}_{A}\hat{P}_{B}|0\rangle
\\
\langle0|\hat{Q}_{A}\hat{Q}_{B}|0\rangle &
 \langle0|\hat{P}_{A}\hat{Q}_{B}|0\rangle & 
 \langle0|\hat{Q}_{B}^{2}|0\rangle & 
 \matrixel{0}{\hat{Q}_B\hat{P}_B}{0}
 \\
\langle0|\hat{Q}_{A}\hat{P}_{B}|0\rangle &
 \langle0|\hat{P}_{A}\hat{P}_{B}|0\rangle &
\matrixel{0}{\hat{Q}_B\hat{P}_B}{0} &
 \langle0|\hat{P}_{B}^{2}|0\rangle
\end{array}
\right) \ , 
\nonumber
\end{align}
a Gaussian quantum state $\hat{\rho}$ of two harmonic oscillators is fixed.
The state $\hat{\rho}$ reproduces the entire covariance matrix elements such as%
\begin{equation}
\operatorname*{Tr}\left[  \hat{\rho}\hat{O}_{A}\hat{O}_{B}^{\prime}\right]  =\langle
0|\hat{O}_{A}\hat{O}_{B}^{\prime}|0\rangle.
\end{equation}
When $\hat{\rho}$ is a \textit{pure state} of the two oscillators, mode $B$ is referred to as the partner of mode $A$. Then the covariance  matrix can take the following simple form:
\begin{equation}
M_{AB}=\left[
\begin{array}
[c]{cccc}%
\frac{1}{2}\sqrt{1+g^{2}} & 0 & \frac{g}{2} & 0\\
0 & \frac{1}{2}\sqrt{1+g^{2}} & 0 & -\frac{g}{2}\\
\frac{g}{2} & 0 & \frac{1}{2}\sqrt{1+g^{2}} & 0\\
0 & -\frac{g}{2} & 0 & \frac{1}{2}\sqrt{1+g^{2}}%
\end{array}
\right]  .\label{4}%
\end{equation}
Its corresponding wave-function of the pure state is given as
\begin{equation}
\langle Q_A,Q_B|\Psi_{AB}\rangle\varpropto\exp\left[  -\frac{1}{2}%
\sqrt{1+g^{2}}\left(  Q_A^{2}+Q_B^{2}\right)  +gQ_AQ_B\right]
,\label{3}%
\end{equation}
in the correlation space \cite{cc1} \cite{cc2}. When the support of the real coefficients $( X_B(n),Y_B(n),Z_B(n),W_B(n))  $ does not have overlap with the support of $(X_A(n) , $ $ Y_A(n), Z_A(n), W_A(n))  $, mode $B$ is referred to as \textit{spatially separated partner }of mode $A$. When the support of the real coefficients $( X_B(n),Y_B(n),Z_B(n),W_B(n))  $  has nonzero overlap with the support of $(X_A(n) , Y_A(n), Z_A(n), W_A(n))  $, mode $B$ is referred to as
\textit{spatially overlapped partner} of mode $A$. Even though mode $A$ overlaps mode $B$, they are locally independent when the following conditions are satisfied:%
\begin{align}
\left[  \hat{Q}_{A},\hat{Q}_{B}\right]   &  =0,\label{1}\\
\left[  \hat{Q}_{A},\hat{P}_{B}\right]   &  =0,\\
\left[  \hat{P}_{A},\hat{Q}_{B}\right]   &  =0,\\
\left[  \hat{P}_{A},\hat{P}_{B}\right]   &  =0. \label{2}%
\end{align}

No operation $U_{B}\left(  \hat{Q}_{B},\hat{P}_{B}\right)  $ on mode $B$ generated by $\left(  \hat{Q}_{B},\hat{P}_{B}\right)  $ affects mode $A$, and no operation $U_{A}\left(  \hat{Q}_{A},\hat{P}_{A}\right)  $ on mode $A$ generated by $\left(  \hat{Q}_{A},\hat{P}_{A}\right)  $ affects mode $B$. Thus, in the correlation space spanned by $\left(  \hat{Q}_{A},\hat{P}_{A},\hat{Q}_{B},\hat{P}_{B}\right)  $ \cite{cc1} \cite{cc2}, $A$ and $B$ are actually locally independent. Since locality of $A$ and $B$ can be introduced in the above way, quantum entanglement among $A$ and $B$ are well defined. From equation (\ref{3}), the entanglement entropy is computed as
\begin{equation}
S_{EE}(A,B)=\sqrt{1+g^{2}}\ln\left(  \frac{1}{g}\left(  \sqrt{1+g^{2}%
}+1\right)  \right)  +\ln\left(  \frac{g}{2}\right)  .
\end{equation}
By solving equation (\ref{4}), as it was done in reference \cite{TYH}, 
 we are able to identify the partner $B$ window functions $(X_B(n), Y_B(n), Z_B(n), W_B(n))$%
\begin{eqnarray}
X_B(n)&=& \frac{\sqrt{1+g^2}}{g} X_A(n) -\frac{2}{g} \sum_{n^\prime=1}^N \Delta_p(n-n^\prime) W_A(n^\prime)
\ , 
\label{XBn}
\\
Y_B(n)&=& \frac{\sqrt{1+g^2}}{g} Y_A(n) +\frac{2}{g} \sum_{n^\prime=1}^N \Delta_q(n-n^\prime) Z_A(n^\prime)
\ , 
\label{YBn}
\\
Z_B(n)&=& - \frac{\sqrt{1+g^2}}{g} Z_A(n) -\frac{2}{g} \sum_{n^\prime=1}^N \Delta_p(n-n^\prime) Y_A(n^\prime)
\ , 
\label{ZBn}
\\
W_B(n)&=&- \frac{\sqrt{1+g^2}}{g} W_A(n) +\frac{2}{g} \sum_{n^\prime=1}^N \Delta_q(n-n^\prime) X_A(n^\prime)
\ ,
\label{WBn}
\end{eqnarray}
 Notice that these equations depend on the original mode $A$ window functions $(X_A(n), Y_A(n), $ $Z_A(n), W_A(n))$ after the symplectic transformation.  The first term to the right hand size in  equations \eqref{XBn}-\eqref{WBn}, the one without the summation, implies that in general the purification partner $B$ has an overlap with the original mode $A$. Only for some specific window functions $(X_A(n), Y_A(n), Z_A(n),$ $ W_A(n))$, with support in a region $1\leq n\leq \tilde{n}$ with $\tilde{n}$ an integer, such as the right hand size  in  equations \eqref{XBn}-\eqref{WBn} vanishes for a region $\tilde{n}<n\leq N$ we have spatial separability between the support of $A$ and $B$.  
\newpage
\section{\bigskip Partner Harvesting}
In this section the entanglement harvesting protocol based on SOP is presented.  Our protocol allow us to to obtain an infinite amount of harvested entanglement by tuning up the window functions of the original mode $A$.   For simplicity let us consider the case in which the original mode $A$ covariance matrix $m_A$ does not contain $q$-$p$ correlations, that is $y_A(n)=z_A(n)=0$ such as the symplectic transformation does not mixed $q$ and $p$:
\begin{align}
\hat{q}_A  &  =\sum_{n=1}^{N}x_A(n)\hat{q}_{n}\ ,
\\
\hat{p}_A  &  =\sum_{n=1}^{N}w_A(n)\hat{p}_{n}\ ,
\end{align}%
where the window function $x_A(n)$ and $w_A$ satisfy the condition imposed by the commutation relation in equation \eqref{comm}
\begin{equation}
\sum_{n=1}^N x_A(n) w_A(n) = 1 \ .
\label{xAwA}
\end{equation}
 Considering a symplectic transformation such as:
\begin{align}
\hat{Q}_{A}  &  =\left(  \frac{\langle 0|\hat{p}_{A}^{2}|0\rangle}{\langle
0|\hat{q}_{A}^{2}|0\rangle}\right)  ^{1/4}\hat{q}_A \ , 
\\
\hat{P}_{A}  &  =\left(  \frac{\langle 0|\hat{p}_{A}^{2}|0\rangle}{\langle
0|\hat{q}_{A}^{2}|0\rangle}\right)  ^{-1/4}\hat{p}_A \ ,
\end{align}
we get
\begin{eqnarray}
M_{A}&=&\left(
\begin{array}
[c]{cc}%
\frac{1}{2}\sqrt{1+g^{2}} & 0\\
0 & \frac{1}{2}\sqrt{1+g^{2}}%
\end{array}
\right) \ .
\end{eqnarray}
For this particular case of no $q$-$p$ correlations the factor $g$ is given by:
\begin{equation}
g=\sqrt{4\langle 0|\hat{Q}_{A}^{2}|0\rangle\langle 0|\hat{P}_{A}^{2}%
|0\rangle-1}=\sqrt{4\langle 0|\hat{q}_A^{2}|0\rangle\langle
0|\hat{p}_A^{2}|0\rangle-1} \ . 
\label{g2}
\end{equation}
By calculating the expectation values we get:
\begin{eqnarray}
\langle 0|\hat{q}_A^{2}|0\rangle&=& \sum_{k=1}^N \sum_{l=1}^N x_A(k) \Delta_q(k-l) x_A(l)  \ ,
\label{qA2}
\\
\langle 0|\hat{p}_A^{2}|0\rangle&=& \sum_{m=1}^N \sum_{n=1}^N w_A(m) \Delta_p(m-n) w_A(n)  \ . 
\label{pA2}
\end{eqnarray}
 Consider the case in which the window functions $x_A(n)$ and $w_A(n)$ are restricted to one specific site $n^\prime$ such as $\hat{q}_A=\hat{q}_{n^\prime}$ and $\hat{p}_A=\hat{p}_{n^\prime}$. For this case, it can be shown that the g factor in equation \eqref{g2} is finite and given by
 \begin{equation}
g= \sqrt{4 \Delta_q(0) \Delta_p(0)-1} \ . 
\end{equation}
In the case the original mode $A$ is restricted to only one site, it is not possible to obtain an infinite amount of entanglement. Notice that in the discrete model a one site only window function is equivalent to a point like function for the continuum model.  That is point like UdW detectors associated to the original mode $A$ cannot harvest an infinite amount of entanglement with our protocol. 
 
 From now on, we consider the original mode $A$ with localized window functions that involve more than one site. Equivalently in the continuum model, an UdW detector associated to the original mode $A$ has some spatial smearing. The tuning of the window functions $x_A(n)$ and $w_A(n)$ allow us to have an infinite amount of entanglement between original mode $A$ and partner $B$.  The window functions must satisfy the constraint coming from the commutation relation in equation \eqref{xAwA}. This guarantees a non-divergent behavior for the sum of product of window functions $x_A(n^\prime) w_A(n^\prime)$ on the same site $n^\prime$. However, the factor $g$ that quantifies the entanglement depends also on products of window function on different sites.  From equations  \eqref{qA2} and \eqref{pA2} we can see that the factor g in equation \eqref{g2} is given by the sum of terms like $x_A(l) \Delta_q(l-k) x_A(k)w_A(m) \Delta_p(m-n) w_A(n)$ for $k, l , m, n$ taking values from $1$ to $N$.  Therefore, by choosing window functions such as \eqref{xAwA} is satisfied while one of the products $x_A(l) x_A(k)w_A(m) w_A(n)$ diverges, it is possible to have an infinite amount of entanglement between the original mode $A$ and its associated purification partner $B$. For example let us consider the original mode $A$ defined by the window functions: 
\begin{equation}
x_A(1)=1, \hspace{1cm} x_A(2)=0, \hspace{1cm}x_A(n)=0  \ \text{for} \ 3\leq n \leq N \ ,
\label{model1}
\end{equation}
\begin{equation}
w_A(1)=1, \hspace{1cm} w_A(2)=\frac{1}{\delta}, \hspace{1cm}w_A(n)=0  \ \text{for} \ 3\leq n \leq N \ , 
\end{equation}
or equivalently 
\begin{align}
\hat{q}_A  &  =\hat{q}_{1} \ , \\
\hat{p}_A  &  =\hat{p}_{1}+\frac{1}{\delta}\hat{p}_{2} \ .  %
\end{align}%
In this model $\delta$ is the parameter that allow us to obtain an infinite amount of entropy in the limit $\delta\rightarrow0$. 
From which the factor $g$ that quantifies the entanglement can be calculated as:
\begin{equation}
\frac{1}{2}\sqrt{1+g^{2}}=\sqrt{\Delta_{q}(0)\left(  \Delta_{p}(0)\left(
1+\frac{1}{\delta^{2}}\right)  +\frac{2}{\delta}\Delta_{p}(1)\right)  }
 \ .
 \label{model2}
\end{equation}
In the small $\delta$ limit the dominant term in $g$ behaves like:
\begin{equation}
g\sim\frac{1}{\delta}  \ . %
\end{equation}%
In the $\delta\rightarrow0$ limit the entanglement entropy diverges 
\begin{equation}
S_{EE}\sim\log{\left(  \frac{1}{\delta}\right)  }\rightarrow \infty \ .%
\label{SN3d0}
\end{equation}
 
  Since for SSP the form of the window function is determined from imposing the separability conditions in equations  \eqref{XBn}-\eqref{WBn}, any tuning of the window functions to obtain an infinite amount of entanglement is related to SOP. In addition, notice that since we are working on a discrete model , the entanglement divergence is not related to the continuum limit of the field theory. In section 4 we present a simplest example with just $N=3$ harmonic oscillators for which we calculate the energy cost to harvest that entanglement.   
  
\begin{figure}[h]
\centering
\includegraphics[scale=1.0]{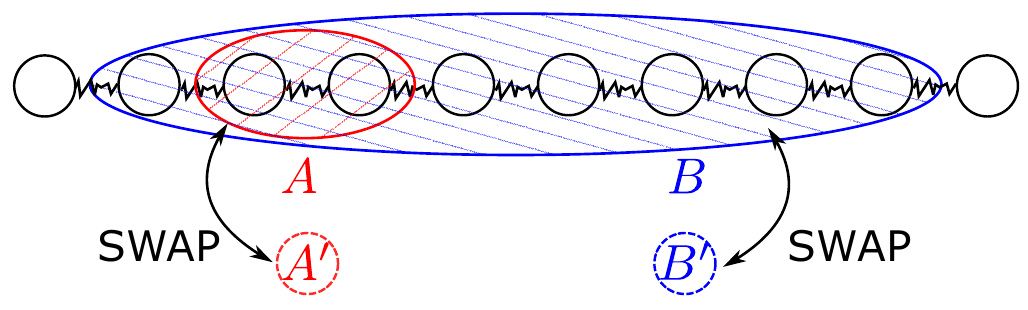}
\caption[Entanglement harvesting setup]{Entanglement harvesting protocol based on SOP.  The entanglement shared between the original mode $A$ and its corresponding overlapped partner $B$ is swapped to two external devices  $A^\prime$ $B^\prime$. The original quantum system from which $A$ and $B$ are constructed is represented here as a 1-dimensional harmonic oscillator chain. }
\label{protocol}
\end{figure}

We proceed now to enumerate the steps in our entanglement harvesting protocol. This one is based on  swapping operations between the mode $A$ and its corresponding partner $B$, with two external devices $A^\prime$ and $B^\prime$. We consider the case in which we instantaneously couple the original system $AB$ to the external devices  $A^\prime B^\prime$, such as we can neglect the dynamics of the latter system. In addition we assume that the original system is initially in the ground state $\ket{0}$ while the external devices are in the state $\ket{\psi_{A^\prime}}$ and $\ket{\psi_{B^\prime}}$ for systems $A^\prime$ and $B^\prime$ respectively. A schematic picture of our entanglement harvesting protocol can be found in figure \ref{protocol}. 

 The steps of our protocol are as follows:
\begin{enumerate}%
\item We consider a swapping operation $\hat{U}_{AA^\prime}$ as follows:
\begin{equation}
\hat{U}_{AA^{\prime}}=\exp\left[  \imath\frac{\pi}{2}\left(  \hat{Q}_A\hat
{p}_{A^{\prime}}-\hat{P}_{A}\hat{q}_{A^{\prime}}\right)  \right] \ . 
\label{UA}
\end{equation}
Notice that this operation can be interpreted as a rotation in the plane $(\hat{Q}_A,\hat{q}_{A^\prime})$. It should be reminded that we have the difference  $\hat{p}_A\neq \hat{p}_{A^\prime}$; the first is an operator defining partner $A$, while the latter corresponds to the external device $A^\prime$.  The operation $\hat{U}_{AA^\prime}$ can be interpreted as a rotation in the $(\hat{q}_{A^\prime}, \hat{Q}_A)$ plane by an angle of $\theta=\pi/2$. 
\begin{align}
\hat{U}_{AA^{\prime}}^{\dagger}\hat{Q}_A\hat{U}_{AA^{\prime}}  &  =\hat
{q}_{A^{\prime}} \ .
\\
\hat{U}_{AA^{\prime}}^{\dagger}\hat{q}_{A^{\prime}}\hat{U}_{AA^{\prime}}  &
=-\hat{Q}_A \ . %
\end{align}

\item Immediately after the first operation we apply a second swapping operation $\hat{U}_{BB^\prime}$ such that
\begin{equation}
\hat{U}_{BB^{\prime}}=\exp\left[  \imath\frac{\pi}{2}\left(  \hat{Q}_B\hat
{p}_{B^{\prime}}-\hat{P}_{B}\hat{q}_{B^{\prime}}\right)  \right] \ .
\label{UB}
\end{equation}
Similarly  $\hat{U}_{BB^{\prime}}$ can be interpreted as a rotation in the $(\hat{q}_{B^\prime}, \hat{Q}_B)$ plane by an angle of $\theta=\pi/2$. 
\begin{align}
\hat{U}_{BB^{\prime}}^{\dagger}\hat{Q}_B\hat{U}_{BB^{\prime}}  &  =\hat
{q}_{B^{\prime}} \ ,
\\
\hat{U}_{BB^{\prime}}^{\dagger}\hat{q}_{B}\hat{U}_{BB^{\prime}}  &  =-\hat{Q}_B  \ . %
\end{align}
After this swapping operation, the total state $\ket{\Psi_{A^{\prime}B^{\prime}}}$ of the external device is an entangled state between subsystem $A^\prime$ and $B^\prime$. 
\begin{equation}
\langle q_{A^{\prime}},q_{B^{\prime}}|\Psi_{A^{\prime}B^{\prime}}%
\rangle\varpropto\exp\left[  -\frac{1}{2}\sqrt{1+g^{2}}\left(  q_{A^{\prime}%
}^{2}+q_{B^{\prime}}^{2}\right)  +gq_{A^{\prime}}q_{B^{\prime}}\right]  .
\end{equation}
\end{enumerate}
The pure state entanglement of the field sub-system composed by the original mode $A$ and its associated purification partner $B$ was harvested (swapped) to two external devices $A^\prime$ and $B^\prime$. By tuning the window functions  $x_A(n)$ and $w_A(n)$, our protocol achieves an infinite amount of entanglement harvesting.  In the next section we apply this protocol to $N=3$ harmonic oscillators and we calculate the energy cost associated to the entanglement extraction. 

\section{\bigskip Energy Cost of Partner Harvesting}

Let us consider a system of $N=3$ three harmonic oscillators described by equations \eqref{model1}-\eqref{model2}. For this model the correlation functions in equations \eqref{corrq} and \eqref{corrp} can be written as
\begin{align}
\Delta_{q}(0)  &  =\frac{1}{6}\left(  \frac{1}{\omega_{0}}+\frac{2}{\omega
_{1}}\right) \ ,
\\
\Delta_{p}(0)  &  =\frac{1}{6}\left(  \omega_{0}+2\omega_{1}\right)  \ , 
\\
\Delta_{q}(1)  &  =\Delta_{q}(2)=\frac{1}{6}\left(  \frac{1}{\omega_{0}}%
-\frac{1}{2}\left(  \frac{1}{\omega_{1}}+\frac{1}{\omega_{2}}\right)  \right)
\ ,
\\
\Delta_{p}(1)  &  =\Delta_{p}(2)=\frac{1}{6}\left(  \omega_{0}-\frac{1}%
{2}\left(  \omega_{1}+\omega_{2}\right)  \right) \ .
\end{align}%
 In order to use the partner formula to find the partner $B$ we need to perform the symplectic transformation. Since there is no mixing between $q$ and $p$, in the symplectic transformation, we have $\theta_A=\theta_A^\prime=0$. The solution for $\sigma_A$ is written:
 \begin{equation}
\exp{\sigma_A}=\left(\frac{\matrixel{0}{\hat{p}_A^2}{0}}{\matrixel{0}{\hat{q}_A^2}{0}} \right)^{1/4} \ . 
 \end{equation}
The original mode $A$ after the symplectic transformation is written as:
\begin{eqnarray}
\hat{Q}_A&=& C \hat{q}_A \ , 
\label{QAN3}
\\
\hat{P}_A&=& \frac{1}{C}  \hat{p}_A \ ,
\label{PAN3}
\end{eqnarray} 
where we have defined:
\begin{eqnarray}
C= \left[\frac{\Delta_p(0) \left(1+\frac{1}{\delta^2}\right) +\frac{2}{\delta} \Delta_p(1)}{\Delta_q(0)}\right]^{1/4} \ .
\end{eqnarray}
For our three oscillator chain, from equations \eqref{QAN3} and \eqref{PAN3} we can identify:
\begin{eqnarray}
X_A(1)&=& \left(\frac{\sqrt{1+g^2}}{2}\right)^{-1/2} C  \ ,
\label{XA1}
\\
W_A(1)&=&  \left(\frac{\sqrt{1+g^2}}{2}\right)^{-1/2}\frac{1}{C}  \ , 
\\
W_A(2)&=& \left(\frac{\sqrt{1+g^2}}{2}\right)^{-1/2} \frac{1}{\delta} \frac{1}{C} \ , 
\label{WA2}
\end{eqnarray} 
with all the other components of $X_A(n), Y_A(n), Z_A(n)$ and $W_A(n)$ equal to zero. After the symplectic transformation mode $A$ can be written as:
\begin{eqnarray}
\hat{Q}_A&=& \left(\frac{\sqrt{1+g^2}}{2}\right)^{1/2} X_A(1) \hat{q}_1 \ , 
\label{N3QA}
\\
\hat{P}_A&=& \left(\frac{\sqrt{1+g^2}}{2}\right)^{1/2} \left(W_A(1) \hat{p}_1 +W_A(2) \hat{p}_2\right) \ . 
\label{N3PA}
\end{eqnarray} 
On the other hand, the partner $B$ can be constructed by substituting equations \eqref{XA1}-\eqref{WA2} in equations \eqref{XBn}-\eqref{WBn}. The results are as follows:
\begin{eqnarray}
X_B(1)&=& \frac{\sqrt{1+g^2}}{g} X_A(1) -\frac{2}{g} \left[\Delta_p(0)W_A(1)+\Delta_p(1)W_A(2)\right]
\ ,
\\
X_B(2)&=& -\frac{2}{g} \left[\Delta_p(1)W_A(1)+\Delta_p(0)W_A(2)\right]
\ ,
\\
X_B(3)&=&  -\frac{2}{g} \left[\Delta_p(2)W_A(1)+\Delta_p(1)W_A(2)\right]
\ , 
\end{eqnarray} 
\begin{eqnarray}
W_B(1)&=& -\frac{\sqrt{1+g^2}}{g} W_A(1) +\frac{2}{g} \Delta_q(0)X_A(1) \ ,
\\
W_B(2)&=&-\frac{\sqrt{1+g^2}}{g} W_A(2) +\frac{2}{g} \Delta_q(1)X_A(1) \ , 
\\
W_B(3)&=& \frac{2}{g} \Delta_q(2)X_A(1) \ ,
\end{eqnarray} 
with all the other components of $Y_B(n)$ and $Z_B(n)$ equal to zero. The partner $B$ for our three oscillator toy model can be written as:
\begin{eqnarray}
\hat{Q}_B&=& \left(\frac{\sqrt{1+g^2}}{2}\right)^{1/2} \left(X_B(1) \hat{q}_1 +X_B(2) \hat{q}_2 +X_B(3) \hat{q}_3 \right) \  , 
\label{QBN3}
\\
\hat{P}_B&=&\left(\frac{\sqrt{1+g^2}}{2}\right)^{1/2} \left(W_B(1) \hat{p}_1 +W_B(2) \hat{p}_2 +W_B(3) \hat{p}_3 \right) \ . 
\label{PBN3}
\end{eqnarray} 

Notice that mode $A$ in equations \eqref{N3QA} and \eqref{N3PA} is localized. It is only composed of operators of oscillators  $1$ and $2$. On the other hand,  partner $B$ is conformed by contributions from all three oscillators. There is an overlap between mode $A$ and partner $B$ window functions. That is, we have a case of SOP. 

\begin{figure}[h]
\centering
\includegraphics[scale=1]{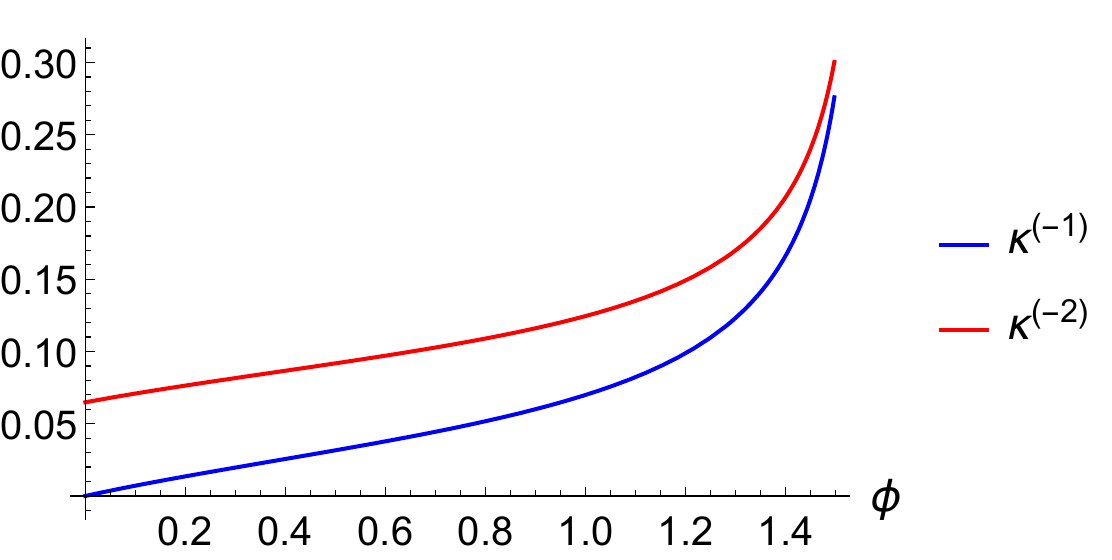}
\caption[Part of Energy cost independent of the external devices]{ Coefficient $\kappa^{(-1)}$ and $\kappa^{(-2)}$  of the Taylor series around $\delta=0$ of the energy cost term $\kappa$ independent of the initial state of the external devices. A change of variables from $\eta \in(0,\infty)$ to $\phi\in(0,\pi/2)$ was done trough $\eta=\tan(\phi)$ in order to study the behavior of the Taylor series coefficients. Notice that for all possible values of $\phi$ the coefficients are always positive. }
\label{divergences}
\end{figure}

\begin{figure}[h]
\centering
\includegraphics[scale=1]{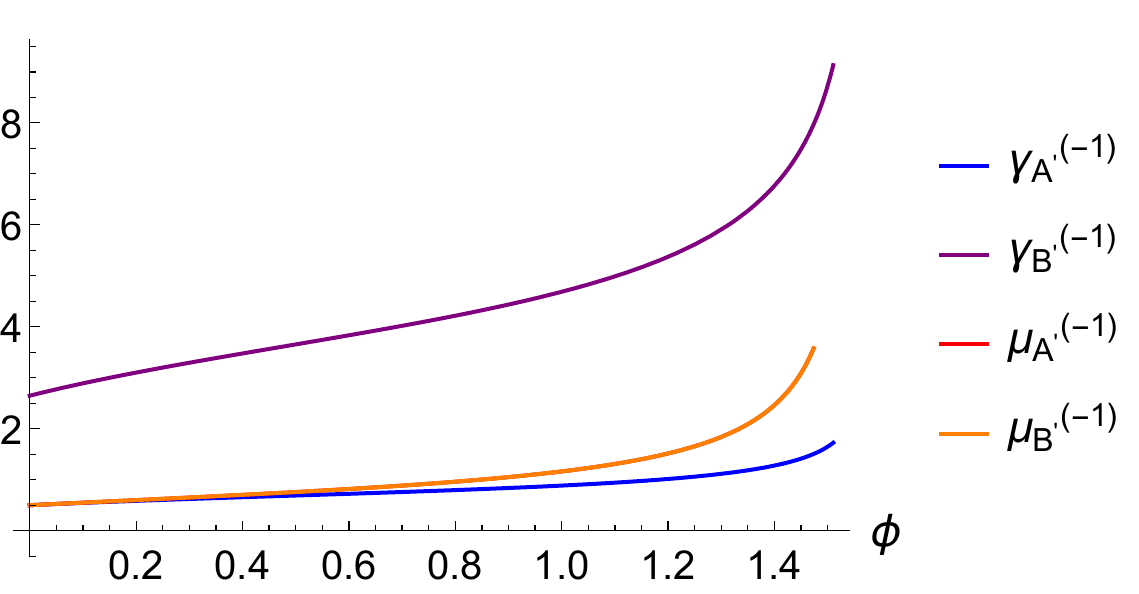}
\caption[Part of the Energy cost that depends on the external devices]{Part of the energy cost that depends on the initial state of the external devices. Coefficient $\delta^{-1}$ of the Taylor series around $\delta=0$ of functions: $\mu_{A^\prime}, \mu_{B^\prime}, \gamma_{A^\prime}, \gamma_{B^\prime}$. The coefficient of $\mu_{A^\prime}^{(-1)}$ is equal to $\mu_{B^\prime}^{(-1)}$. A change of variables from $\eta \in(0,\infty)$ to $\phi\in(0,\pi/2)$ was done trough $\eta=\tan(\phi)$ in order to study the behavior of the Taylor series coefficients. Notice that for all possible values of $\phi$ the coefficients are always positive. }
\label{optimizes}
\end{figure}

We now apply the entanglement harvesting protocol of section 3 to the original mode in equations \eqref{N3QA} \eqref{N3PA} and the corresponding partner in equations  \eqref{QBN3} \eqref{PBN3}. The original system of the three oscillators is in the ground state $\left\vert 0\right\rangle $ while the external devices are in the state$\left\vert \psi_{A^{\prime}}\right\rangle $ and $\left\vert \psi_{B^{\prime}}\right\rangle $ for systems $A^{\prime}$ and $B^{\prime}$ respectively. Let us first focus on the case where the Hamiltonian of the external devices is zero. The energy cost to do our swapping operation $\Delta E_{swap}$ can be calculated as:
\begin{equation}
\Delta E_{swap}=\left\langle \hat{H}_{AA^{\prime}BB^{\prime}}\right\rangle
-\left\langle \hat{H}\right\rangle \ , 
\end{equation}
where we have defined
\begin{equation}
\left\langle \hat{H}_{AA^{\prime}BB^{\prime}}\right\rangle \equiv\left\langle
0\right\vert _{AB}\left\langle \psi_{A^{\prime}}\right\vert \left\langle
\psi_{B^{\prime}}\right\vert \hat{U}_{BB^{\prime}}^{\dagger}\hat
{U}_{AA^{\prime}}^{\dagger}\hat{H}\hat{U}_{AA^{\prime}}\hat{U}_{BB^{\prime}%
}\left\vert 0\right\rangle _{AB}\left\vert \psi_{A^{\prime}}\right\rangle
\left\vert \psi_{B^{\prime}}\right\rangle 
\end{equation}
and
\begin{equation}
\left\langle \hat{H}\right\rangle \equiv\left\langle 0\right\vert _{AB}\left\langle
\psi_{A^{\prime}}\right\vert \left\langle \psi_{B^{\prime}}\right\vert
\hat{H}\left\vert 0\right\rangle _{AB}\left\vert \psi_{A^{\prime}}\right\rangle
\left\vert \psi_{B^{\prime}}\right\rangle \ .
\end{equation}
The details of the calculation of the energy cost can be found in appendix \ref{appendix}. The results for the energy cost by the swapping operation are as follows. The average energy of the three oscillators in the ground state $\avg{\hat{H}}$ is independent of the initial state of the external devices. Let us focus on the remaining  term:
\begin{eqnarray}
\avg{\hat{H}_{AA^\prime BB^\prime}}&=& \alpha_p \Delta_p(0) +\beta_p \Delta_p(1) +\alpha_q \Delta_q(0) +\beta_q \Delta_q(1)
\nonumber
\\
&& + \gamma_{A^\prime} \avg{\hat{p}_{A^\prime}^2} + \mu_{A^\prime} \avg{\hat{q}_{A^\prime}^2} +\gamma_{B^\prime} \avg{\hat{p}_{B^\prime}^2} +\mu_{B^\prime} \avg{\hat{q}_{B^\prime}^2} \ , 
\label{cost}
\end{eqnarray}
where the coefficients $\alpha_p, \beta_p, \alpha_q, \beta_q, \gamma_{A^\prime}, \mu_{A^\prime}, \gamma_{B^\prime}$ and $\mu_{B^\prime}$ are all long expressions that can be found in appendix  \ref{appendix}. The remaining expectation values  $\avg{\hat{q}_{A^\prime}^2},  \avg{\hat{p}_{A^\prime}^2}$ and  $\avg{\hat{q}_{B^\prime}^2}, \avg{\hat{p}_{B^\prime}^2}$ are with respect to the initial states of the external devices $\ket{\psi_{A^\prime}}$ and $\ket{\psi_{B^\prime}}$ respectively.  The first line in equation \eqref{cost} contains all the terms that are independent of the initial state of the external devices $A^\prime B^\prime$. Let us introduce a new label $\kappa$ for all of those contributions
\begin{equation} 
\kappa = \alpha_p \Delta_p(0) +\beta_p \Delta_p(1) +\alpha_q \Delta_q(0) +\beta_q \Delta_q(1) \ . 
\end{equation}
This term can be expanded in a Taylor series around $\delta=0$
\begin{equation}
\kappa= \frac{1}{\delta^2} \kappa^{(-2)} + \frac{1}{\delta} \kappa^{(-1)} + \kappa^{(0)} + \mathcal{O}(\delta) \ ,
\end{equation}
where the coefficients $\kappa^{(-2)} $ et al are functions of $\eta$ the coupling constant. Since this constant can take positive values between $(0,\infty)$, by introducing a change of variables $\eta=\tan(\phi)$ with $\phi\in(0, \pi/2)$, it is possible to study the behavior of  the dominant terms $\kappa^{(-2)}$ and $\kappa^{(-1)}$ of the Taylor series around $\delta=0$.  The coefficients as a function of $\phi$ can be seen in figure \ref{divergences}. Both coefficients  $\kappa^{(-2)}$ and $\kappa^{(-1)}$ are positive for all possible  values of $\eta\in (0,\infty)$.  Similarly, each of the terms in the second  line in equation \eqref{cost}, the coefficients that come together with expectation values with respect to the initial state of the external devices, can also be expanded in a Taylor series around $\delta=0$ as follows:
\begin{eqnarray}
\gamma_{A^\prime}&=& \frac{1}{\delta} \gamma_{A^\prime}^{(-1)} + \gamma_{A^\prime}^{(0)}+ \mathcal{O}(\delta) \ , 
 \\
 \mu_{A^\prime}&=& \frac{1}{\delta} \mu_{A^\prime}^{(-1)}+ \mu_{A^\prime}^{(0)}+ \mathcal{O}(\delta) \ , 
 \\
 \gamma_{B^\prime}&=& \frac{1}{\delta} \gamma_{B^\prime}^{(-1)} + \gamma_{B^\prime}^{(0)} + \mathcal{O}(\delta) \ , 
 \\
 \mu_{B^\prime}&=& \frac{1}{\delta} \mu_{B^\prime}^{(-1)} + \mu_{B^\prime}^{(0)} + \mathcal{O}(\delta) \ . 
\end{eqnarray}
By introducing the same change of variables as before,  $\eta=\tan(\phi)$ with $\phi\in(0, \pi/2)$ the behavior of the Taylor series coefficients associated to $\delta^{-1}$ can be seen in figure \ref{optimizes}. 

The divergent part of $\mu_{A^\prime}, \mu_{B^\prime}, \gamma_{A^\prime}, \gamma_{B^\prime}$ when $\delta\rightarrow0$ is positive for all possible values of $\eta\in (0,\infty)$ such as the Hamiltonian in equation \eqref{H} represents a physical system of coupled harmonic oscillators.  In addition, the expectation values: $\avg{\hat{q}^2_{A^\prime}}$, $\avg{\hat{p}^2_{A^\prime}}$, $\avg{\hat{q}^2_{B^\prime}}$ and$\avg{\hat{p}^2_{B^\prime}}$ are always positive. Therefore even through optimization of the external devices initial states $\ket{\psi_{A^\prime}(\delta)}$ and $\ket{\psi_{B^\prime}(\delta)}$ it is impossible to cancel out the divergences in equation \eqref{cost}.  We conclude then that the energy cost for this entanglement swapping toy model behaves like:
 \begin{equation}
 \Delta E_{swap}\sim \frac{1}{\delta^2} , 
\label{Ecostd1}
 \end{equation}
  which implies, in the limit of $\delta\rightarrow0$,  the energy cost to extract an infinite amount of pure state entanglement from the purification partners to an external system is infinite  
  \begin{equation}
 \lim_{\delta\rightarrow 0}\Delta E_{swap} \rightarrow \infty.
\label{Ecostd0}
 \end{equation}
  With the swapping operation in this toy model of just three oscillators it is possible to harvest an infinite amount of entanglement as predicted in equation \eqref{SN3d0}, but an infinite amount of energy \eqref{Ecostd0} is needed to harvest that entanglement. 
 
Let $\Delta E$ be the energy cost when we have non-vanishing Hamiltonians $\hat{H}_{A^\prime}$ and $\hat{H}_{B^\prime}$ of the external devices and the initial states $\ket{\psi_{A^\prime}}$ and $\ket{\psi_{B^\prime}}$ correspond to the ground states of each external device. This energy cost $\Delta E$ is lower bounded by the energy cost $\Delta E_{swap}$ which we just calculated in equation  \eqref{Ecostd1}.  Therefore, the energy cost $\Delta E$ also diverges when $\delta\rightarrow 0$.

\section{Summary and Discussion}

We proposed a new entanglement harvesting protocol based on spatially overlapped purification partners. This new protocol allow us to extract a huge amount of  pure state entanglement from two subsystems of a quantum field. This infinite amount of entanglement extraction is attained without taking the continuum limit of the field theory. For a three coupled harmonic oscillators model we explicitly calculate the energy cost associated to a huge entanglement extraction. For this model, the energy cost diverges when considering the infinite entanglement extraction limit. It would be interesting to look for  similar trade off relations between extracted entanglement and energy cost for general entanglement harvesting protocols.

\bigskip

\textit{Acknowlegement.- }We would like to thank Achim Kempf and Naoki Watamura for his useful discussions. This research was partially supported by JSPS KAKENHI Grant Numbers 16K05311 (M.H.) and 18J20057 (K.Y.), and by Graduate Program on Physics for the Universe of Tohoku University (K.Y.).
\bigskip

\appendix

\section{Energy cost detailed calculation }
\label{appendix}

\label{details}
In this appendix the detailed calculation concerning the energy cost to swap the entanglement from the three harmonic oscillator system in section 4 is presented. First we want to calculate: $\hat{U}^\dagger_{AA^\prime} \hat{H}\hat{U}_{AA^\prime}$.  Let us consider first an unitary operation depending on a $\theta$ parameter:
\begin{equation}
\hat{U}(\theta)=\exp\left[\imath \theta \hat{G}\right] \ , 
\end{equation}
where $\hat{G}$ is an operator independent of $\theta$. Do not confuse this $\theta$ with a symplectic transformation. After the general calculation we will consider our case of interest $\theta=\pi/2$. Consider an operator $\hat{q}$ such as:
\begin{eqnarray}
\hat{q}(\theta)&=&\hat{U}^\dagger(\theta) \hat{q} \hat{U}(\theta)
\ , 
\\
\partial_\theta \hat{q}(\theta)&=& \imath \commute{\hat{q}(\theta)}{\hat{G}(\theta)} 
\ . 
\end{eqnarray}
For the operator defined in equation \eqref{UA} with arbitrary  $\theta$  
\begin{eqnarray}
U_{AA^\prime}(\theta)=\exp\left[\imath \theta \left(C \hat{q}_1 \hat{p}_{A^\prime}-\frac{1}{C} \left(\hat{p}_1+\frac{1}{\delta} \hat{p}_2\right)\hat{q}_{A^\prime}\right) \right]
\ . 
\end{eqnarray}
From which we can define:
\begin{eqnarray}
\hat{q}_1(\theta)&=&\hat{U}^\dagger_{AA^\prime}(\theta) \hat{q}_1 \hat{U}_{AA^\prime}(\theta)
 \ ,
\\
\hat{q}_2(\theta)&=&\hat{U}^\dagger_{AA^\prime}(\theta) \hat{q}_2 \hat{U}_{AA^\prime}(\theta)
\ ,
\\
\hat{p}_1(\theta)&=&\hat{U}^\dagger_{AA^\prime}(\theta) \hat{p}_1 \hat{U}_{AA^\prime}(\theta)
\ ,
\\
\hat{p}_2(\theta)&=&\hat{U}^\dagger_{AA^\prime}(\theta) \hat{p}_2 \hat{U}_{AA^\prime}(\theta)
\ ,
\\
\hat{q}_{A^\prime}(\theta)&=&\hat{U}^\dagger_{AA^\prime}(\theta) \hat{q}_{A^\prime} \hat{U}_{AA^\prime}(\theta)
\ ,
\\
\hat{p}_{A^\prime}(\theta)&=&\hat{U}^\dagger_{AA^\prime}(\theta) \hat{p}_{A^\prime} \hat{U}_{AA^\prime}(\theta)
\ .
\end{eqnarray}
A short  calculation gives:
\begin{eqnarray}
\partial_\theta \hat{q}_1(\theta)&=& \frac{1}{C} \hat{q}_{A^\prime}(\theta)
\ , 
\\
\frac{1}{C} \partial_\theta \hat{q}_{A^\prime}&=& -\hat{q}_1(\theta) \ .
\end{eqnarray}
By taking second derivatives of each equation it is possible to see that the solutions for  $\hat{q}_1$ and  $\hat{q}_{A^\prime}$ correspond to a harmonic oscillator with unit natural frequency:
\begin{eqnarray}
\partial^2_\theta \hat{q}_1(\theta)&=& - \hat{q}_1(\theta)
 \ ,
\\
\frac{1}{C} \partial^2_\theta \hat{q}_{A^\prime}(\theta)&=& - \frac{1}{C} \hat{q}_{A^\prime}(\theta)
 \ , 
\end{eqnarray}
 with solution in terms of $\theta$
\begin{eqnarray}
\begin{pmatrix}
\hat{q}_1(\theta)
\\
\frac{1}{C} \hat{q}_{A^\prime}(\theta)
\end{pmatrix}
&=&
\begin{pmatrix}
\cos\theta & \sin\theta
\\
-\sin\theta & \cos\theta
\end{pmatrix}
\begin{pmatrix}
\hat{q}_1
\\
\frac{1}{C} \hat{q}_{A^\prime}
\end{pmatrix}
\label{system}
\ .
\end{eqnarray}
From which the desired result is obtained when $\theta=\pi/2$. For $\hat{q}_2$ we have
\begin{equation}
\partial_\theta \hat{q}_2(\theta)= \frac{1}{\delta} \frac{1}{C} \hat{q}_{A^\prime} (\theta) \ ,
\end{equation}
from which the solution can be obtained 
\begin{equation}
\hat{q}_2(\theta)=\hat{q}_2 -\frac{1}{\delta} \hat{q}_1 +\frac{1}{C} \frac{1}{\delta} \left[C \hat{q}_1 \cos\theta +\hat{q}_{A^\prime} \sin\theta\right] 
\ .
\end{equation}
For $\hat{p}_2$ we have no $\theta$ dependence. 
\begin{equation}
\partial_\theta \hat{p}_2(\theta)= 0 
\ ,
\end{equation}
from which the solution can be obtained 
\begin{equation}
\hat{p}_2(\theta)=\hat{p}_2 \ .
\end{equation}
Similarly for $\hat{p}_1$ and $\hat{p}_{A^\prime}$ we have
\begin{eqnarray}
\partial_\theta \hat{p}_1(\theta)&=& C \hat{p}_{A^\prime}(\theta)
\ ,
\\
\partial_\theta \hat{p}_{A^\prime}&=& -\frac{1}{C} \left(\hat{p}_1(\theta) +\frac{1}{\delta} \hat{p}_2(\theta) \right)
\ . 
\end{eqnarray}
By adding the differential equation satisfied by $\hat{p}_2$ then:
\begin{eqnarray}
\partial_\theta \left(\hat{p}_1(\theta)+\frac{1}{\delta} \hat{p}_2\right)&=& C \hat{p}_{A^\prime}(\theta)
\ , 
\\
 C \partial_\theta \hat{p}_{A^\prime}&=& -\left(\hat{p}_1(\theta) +\frac{1}{\delta} \hat{p}_2(\theta) \right)
\ .
\end{eqnarray}
This system of differential equations have a similar solution like the one in equation \eqref{system}. 
\begin{eqnarray}
\hat{p}_1(\theta) +\frac{1}{\delta} \hat{p}_2&=& \left(\hat{p}_1 +\frac{1}{\delta} \hat{p}_2 \right)\cos\theta +C \hat{p}_{A^\prime} \sin\theta
\ . 
\\
 C \hat{p}_{A^\prime}(\theta)&=& - \left(\hat{p}_1+\frac{1}{\delta} \hat{p}_2 \right) \sin\theta + C \hat{p}_{A^\prime} \cos\theta
 \ . 
\end{eqnarray}
 By taking the limit $\theta=\pi/2$:
\begin{eqnarray}
\hat{q}_1\left(\frac{\pi}{2}\right)&=& \frac{1}{C} \hat{q}_{A^\prime}
\ , 
\\
\hat{q}_2\left(\frac{\pi}{2}\right)&=& \hat{q}_2- \frac{1}{\delta} \hat{q}_1 +\frac{1}{C} \frac{1}{\delta} \hat{q}_{A^\prime}
 \ ,
\\
\hat{p}_1\left(\frac{\pi}{2}\right)&=& -\frac{1}{\delta} \hat{p}_2 + C \hat{p}^{A^\prime}
\ ,
\\
\hat{p}_2\left(\frac{\pi}{2}\right)&=&\hat{p}_2
\ .
\end{eqnarray}
These equations together with equation \eqref{H} for $N=3$ allow us to calculate
\begin{eqnarray}
\hat{U}^\dagger_{AA^\prime}\hat{H}\hat{U}_{AA^\prime} &=& 
\frac{1}{2} \left(\left(-\frac{1}{\delta} \hat{p}_2 + C \hat{p}_{A^\prime}\right)^2 +\hat{p}_2^2 +\hat{p}^2_3 \right)
\nonumber
\\ && 
+ \left(\frac{1}{2} +\eta \right)
\left(\frac{1}{C^2} \hat{q}^2_{A^\prime} +\left(\hat{q}_2 -\frac{1}{\delta} \hat{q}_1 +\frac{1}{C} \frac{1}{\delta} \hat{q}_{A^\prime}\right)^2 +\hat{q}_3^2 \right)
\nonumber
\\&&
-\eta \left(\left(\frac{1}{C} \hat{q}_{A^\prime}+\hat{q}_3\right)\left(\hat{q}_2- \frac{1}{\delta} \hat{q}_1 +\frac{1}{C} \frac{1}{\delta} \hat{q}_{A^\prime}\right) +\frac{1}{C} \hat{q}_{A^\prime} \hat{q}_3\right)
\ .
\label{UHU}
\end{eqnarray}

Now we proceed to calculate the effect of the unitary operation $\hat{U}_{BB^\prime}$ over the last equation. Due to the form of equations \eqref{QBN3} and \eqref{PBN3} we have to consider for arbitrary   $\theta$ the operator in equation \eqref{UB} :
\begin{equation}
\hat{U}_{BB^\prime}(\theta)=\exp\left[\imath\theta\left(\hat{Q}_B \hat{p}_{B^\prime} -\hat{P}_B\hat{q}_{B^\prime}\right)\right]
 \ . 
\end{equation}
From which we can define:
\begin{eqnarray}
\hat{q}_1(\theta)&=&\hat{U}^\dagger_{BB^\prime}(\theta) \hat{q}_1 \hat{U}_{BB^\prime}(\theta)
\ ,
\\
\hat{q}_2(\theta)&=&\hat{U}^\dagger_{BB^\prime}(\theta) \hat{q}_2 \hat{U}_{BB^\prime}(\theta)
\ ,
\\
\hat{q}_3(\theta)&=&\hat{U}^\dagger_{BB^\prime}(\theta) \hat{q}_3 \hat{U}_{BB^\prime}(\theta)
\ ,
\\
\hat{p}_1(\theta)&=&\hat{U}^\dagger_{BB^\prime}(\theta) \hat{p}_1 \hat{U}_{BB^\prime}(\theta)
\ ,
\\
\hat{p}_2(\theta)&=&\hat{U}^\dagger_{BB^\prime}(\theta) \hat{p}_2 \hat{U}_{BB^\prime}(\theta)
\ ,
\\
\hat{p}_3(\theta)&=&\hat{U}^\dagger_{BB^\prime}(\theta) \hat{p}_3 \hat{U}_{BB^\prime}(\theta)
\ .
\end{eqnarray}
A short calculations gives the following:
\begin{eqnarray}
\partial_\theta \hat{q}_1 (\theta)&=&  W_B(1) \hat{q}_{B^\prime}(\theta)
 \ ,
\label{j1}
\\
\partial_\theta \hat{q}_2(\theta)&=&  W_B(2) \hat{q}_{B^\prime}(\theta)
\ ,
\\
\partial_\theta \hat{q}_3 (\theta)&=&  W_B(3) \hat{q}_{B^\prime}(\theta)
\ .
\label{j3}
\end{eqnarray}
For $\hat{q}_{B^\prime}(\theta)$ we have:
\begin{equation}
\partial_\theta^2 \hat{q}_{B^\prime}(\theta)= -\Omega^2 \hat{q}_{B^\prime}(\theta)
\ ,
\end{equation}
where we have defined
\begin{equation}
\Omega^2=X_B(1) W_B(1) +X_B(2) W_B(2) +X_B(3) W_B(3) \ .
\end{equation}
 This equation has a solution:
\begin{equation}
\hat{q}_{B^\prime}(\theta)= \hat{q}_{B^\prime} \cos\left(\Omega \theta\right) + \frac{1}{\Omega} \sin\left(\Omega\theta\right) \left[-X_B(1) \hat{q}_1-X_B(2) \hat{q}_2-X_B(3) \hat{q}_3\right] \ . 
\end{equation}
Using this solution we have to solve equations \eqref{j1}-\eqref{j3} summarized as follows:
\begin{equation}
\partial_\theta \hat{q}_j (\theta)=  W_B(j) \hat{q}_{B^\prime}(\theta) \ ,
\end{equation}
for $j=1,2,3$.  Solving the equations:
\begin{equation}
\hat{q}_j(\theta)=\hat{q}_j + W_B(j) \properint{0}{\theta}{q_{B^\prime}(\theta^\prime)}{\theta^\prime} \ .  
\end{equation}
Where the integral term can be calculated:
\begin{align}
\properint{0}{\theta}{\hat{q}_{B^\prime}(\theta^\prime)}{\theta^\prime}&=& \hat{q}_{B^\prime} \frac{1}{\Omega} \sin\left(\Omega \theta\right) + \frac{1}{\Omega^2} \left[1-\cos\left(\Omega\theta\right)\right] \left[-X_B(1) \hat{q}_1-X_B(2) \hat{q}_2-X_B(3) \hat{q}_3\right]
\ . 
\end{align}
With the definitions:
\begin{eqnarray}
d_1&=& -\frac{1}{\Omega^2}\left[1-\cos{\left(\frac{\pi}{2} \Omega\right)}\right] X_B(1)
\ ,
\\
d_2&=& -\frac{1}{\Omega^2}\left[1-\cos{\left(\frac{\pi}{2} \Omega\right)}\right] X_B(2)
\ ,
\\
d_3&=& -\frac{1}{\Omega^2}\left[1-\cos{\left(\frac{\pi}{2} \Omega\right)}\right] X_B(3)
\ ,
\\
d_{B^\prime}&=& \frac{1}{\Omega} \sin{\left(\frac{\pi}{2} \Omega\right)}
\ ,
\end{eqnarray}
for $\theta=\pi/2$ we can write:
\begin{equation}
\properint{0}{\pi/2}{\hat{q}_{B^\prime}(\theta^\prime)}{\theta^\prime}= d_{B^\prime} \hat{q}_{B^\prime} + d_1 \hat{q}_1 +d_2 \hat{q}_2+ d_3 \hat{q}_3
\ . 
\end{equation}
From which:
\begin{equation}
\hat{q}_j\left(\frac{\pi}{2}\right)=\hat{q}_j + W_B(j) \left[d_{B^\prime} \hat{q}_{B^\prime} + d_1 \hat{q}_1 +d_2 \hat{q}_2+ d_3 \hat{q}_3\right]
\ . 
\end{equation}
A similar calculation for $\hat{p}_j$, with $j=1,2,3$ shows:
\begin{equation}
\partial_\theta \hat{p}_j(\theta)= X_B(j) \hat{p}_{B^\prime}(\theta)
\ ,
\end{equation}
\begin{equation}
\partial_\theta \hat{p}_{B^\prime}= -W_B(1) \hat{p}_1(\theta)-W_B(2) \hat{p}_2(\theta)-W_B(3) \hat{p}_3(\theta)
\ . 
\end{equation}
The calculation is similar to the one for $\hat{q}_j$, we just have to exchange:
\begin{eqnarray}
\hat{q}_j&\longleftrightarrow & \hat{p}_j 
\ , 
\\
X_B(j)&\longleftrightarrow & W_B(j)
 \ . 
\end{eqnarray}
The results can be immediately written as:
\begin{equation}
\hat{p}_{B^\prime}(\theta)= \hat{p}_{B^\prime} \cos\left(\Omega \theta\right) + \frac{1}{\Omega} \sin\left(\Omega\theta\right) \left[-W_B(1) \hat{p}_1-W_B(2) \hat{p}_2-W_B(3) \hat{p}_3\right]
 \ ,
\end{equation}
\begin{equation}
\hat{p}_j(\theta)=\hat{p}_j + X_B(j) \properint{0}{\theta}{\hat{p}_{B^\prime}(\theta^\prime)}{\theta^\prime} 
\ , 
\end{equation}
where the integral term can be written 
\begin{align}
\properint{0}{\theta}{\hat{p}_{B^\prime}(\theta^\prime)}{\theta^\prime}&=& \hat{p}_{B^\prime} \frac{1}{\Omega} \sin\left(\Omega \theta\right) + \frac{1}{\Omega^2} \left[1-\cos\left(\Omega\theta\right)\right] \left[-W_B(1) \hat{p}_1-W_B(2) \hat{p}_2-W_B(3) \hat{p}_3\right]
\ . 
\end{align}
With the definitions:
\begin{eqnarray}
s_1&=& -\frac{1}{\Omega^2}\left[1-\cos{\left(\frac{\pi}{2} \Omega\right)}\right] W_B(1)
\ ,
\\
s_2&=& -\frac{1}{\Omega^2}\left[1-\cos{\left(\frac{\pi}{2} \Omega\right)}\right] W_B(2)
\ ,
\\
s_3&=& -\frac{1}{\Omega^2}\left[1-\cos{\left(\frac{\pi}{2} \Omega\right)}\right] W_B(3)
\ ,
\\
s_{B^\prime}&=& d_{B^\prime} =\frac{1}{\Omega} \sin{\left(\frac{\pi}{2} \Omega\right)}
\ ,
\end{eqnarray}
for $\theta=\pi/2$ we can write:
\begin{equation}
\properint{0}{\pi/2}{\hat{p}_{B^\prime}(\theta^\prime)}{\theta^\prime}= s_{B^\prime} \hat{p}_{B^\prime} + s_1 \hat{p}_1 +s_2 \hat{p}_2+ s_3 \hat{p}_3
\ , 
\end{equation}
such as
\begin{equation}
\hat{p}_j\left(\frac{\pi}{2}\right)=\hat{p}_j + X_B(j) \left[s_{B^\prime} \hat{p}_{B^\prime} + s_1 \hat{p}_1 +s_2 \hat{p}_2+ s_3 \hat{p}_3\right]
\ . 
\end{equation}
The last equations allow us to calculate from equation \eqref{UHU}:
\begin{equation}
\hat{U}^\dagger_{BB^\prime}\hat{U}^\dagger_{AA^\prime}\hat{H}\hat{U}_{AA^\prime} \hat{U}_{BB^\prime} = u_1 + u_2 + u_3 + u_4 + u_5 + u_6 + u_7 + u_8 \ , 
\label{UUHUU}
 \end{equation}
 where we defined:
\begin{eqnarray}
u_1&=& \frac{1}{2} \left(-\frac{1}{\delta} \left(\hat{p}_2 + X_B(2) \left[s_{B^\prime} \hat{p}_{B^\prime} + s_1 \hat{p}_1 +s_2 \hat{p}_2+ s_3 \hat{p}_3\right]\right) + C \hat{p}_{A^\prime}\right)^2 
\ , 
\\
u_2 &=& 
\frac{1}{2} \left(\hat{p}_2 + X_B(2) \left(s_{B^\prime} \hat{p}_{B^\prime} + s_1 \hat{p}_1 +s_2 \hat{p}_2+ s_3 \hat{p}_3\right)\right)^2
 \ , 
\\
u_3 &=& 
\frac{1}{2} \left(\hat{p}_3 + X_B(3) \left[s_{B^\prime} \hat{p}_{B^\prime} + s_1 \hat{p}_1 +s_2 \hat{p}_2+ s_3 \hat{p}_3\right]\right)^2
\ , 
\\
u_4&=& \left(\frac{1}{2} +\eta \right)\frac{1}{C^2} \hat{q}^2_{A^\prime}
\ , 
\end{eqnarray}
\begin{eqnarray}
u_5 &=&  \left(\frac{1}{2} +\eta \right)\Bigg(\hat{q}_2 + W_B(2) \left[d_{B^\prime} \hat{q}_{B^\prime} + d_1 \hat{q}_1 +d_2 \hat{q}_2+ d_3 \hat{q}_3\right] 
\nonumber
\\ &&
-\frac{1}{\delta} \left(\hat{q}_1 + W_B(1) \left[d_{B^\prime} \hat{q}_{B^\prime} + d_1 \hat{q}_1 +d_2 \hat{q}_2+ d_3 \hat{q}_3\right]\right) +\frac{1}{C} \frac{1}{\delta} \hat{q}_{A^\prime}\Bigg)^2
\ ,
\\
u_6 &=& \left(\frac{1}{2} +\eta \right)
\left(\hat{q}_3 + W_B(3) \left[d_{B^\prime} \hat{q}_{B^\prime} + d_1 \hat{q}_1 +d_2 \hat{q}_2+ d_3 \hat{q}_3\right]\right)^2
\ ,
\\
u_7&=& 
-\eta \left(\frac{1}{C} \hat{q}_{A^\prime}+\hat{q}_3 + W_B(3) \left[d_{B^\prime} \hat{q}_{B^\prime} + d_1 \hat{q}_1 +d_2 \hat{q}_2+ d_3 \hat{q}_3\right]\right) \times
\nonumber
\\ &&
\times 
\Bigg(\hat{q}_2+ W_B(2) \left[d_{B^\prime} \hat{q}_{B^\prime} + d_1 \hat{q}_1 +d_2 \hat{q}_2+ d_3 \hat{q}_3\right]
\nonumber
\\ &&
- \frac{1}{\delta} \left(\hat{q}_1 + W_B(1) \left[d_{B^\prime} \hat{q}_{B^\prime} + d_1 \hat{q}_1 +d_2 \hat{q}_2+ d_3 \hat{q}_3\right]\right)  +\frac{1}{C} \frac{1}{\delta} \hat{q}_{A^\prime}\Bigg)
\ ,
\\
u_8& =&
-\eta \frac{1}{C} \hat{q}_{A^\prime} \left(\hat{q}_3 + W_B(3) \left[d_{B^\prime} \hat{q}_{B^\prime} + d_1 \hat{q}_1 +d_2 \hat{q}_2+ d_3 \hat{q}_3\right]\right)
\ . 
\end{eqnarray}

Expanding equation \eqref{UUHUU} and taking the expectation value with respect to the ground state of the original system of the three oscillators $\ket{0}$, any term that is linear in terms of the operators $\hat{q}_j$ or $\hat{p}_j$ will vanish since:
\begin{eqnarray}
\matrixel{0}{\hat{q}_j}{0} &=&0 \ , 
\\
\matrixel{0}{\hat{p}_j}{0} &=&0 \ . 
\end{eqnarray}
Furthermore,  any vacuum expectation value of a bi-linear term of the operators $\hat{q}_j$ or  $\hat{p}_j$ will satisfy:
\begin{align}
&&\matrixel{0}{\left(a_1 \hat{q}_1+a_2 \hat{q}_2+a_3 \hat{q}_3\right)\left(b_1 \hat{q}_1+b_2 \hat{q}_2+b_3 \hat{q}_3\right)}{0}= \Delta_q(0) \left(a_1 b_1 +a_2 b_2 + a_3 b_3\right)
\nonumber
\\
 &&
+\Delta_q(1) \left(a_1 b_2 + a_1 b_3 + a_2 b_1 + a_2 b_3 + a_3 b_1 +a_3 b_2\right)
\ , 
\end{align}
 \begin{align}
&&\matrixel{0}{\left(a_1 \hat{p}_1+a_2 \hat{p}_2+a_3 \hat{p}_3\right)\left(b_1 \hat{p}_1+b_2 \hat{p}_2+b_3 \hat{p}_3\right)}{0}= \Delta_p(0) \left(a_1 b_1 +a_2 b_2 + a_3 b_3\right)
\nonumber
\\
&&
+\Delta_p(1) \left(a_1 b_2 + a_1 b_3 + a_2 b_1 + a_2 b_3 + a_3 b_1 +a_3 b_2\right)
\ .
\end{align}
  By considering the previous equations and the results in equation \eqref{UUHUU} it is possible to calculate the part depending on the initial state of the external devices of the energy cost. After collecting terms we end up with equation \eqref{cost}:
\begin{eqnarray}
\avg{\hat{H}_{AA^\prime BB^\prime}}&=& \alpha_p \Delta_p(0) +\beta_p \Delta_p(1) +\alpha_q \Delta_q(0) +\beta_q \Delta_q(1)
\nonumber
\\
&& + \gamma_{A^\prime} \avg{p_{A^\prime}^2} + \mu_{A^\prime} \avg{q_{A^\prime}^2} +\gamma_{B^\prime} \avg{p_{B^\prime}^2} +\mu_{B^\prime} \avg{q_{B^\prime}^2} \ , 
\end{eqnarray}
where we defined:
\begin{eqnarray}
\alpha_p&=&\frac{1}{2 \delta ^2} \left[X_B(2)^2 \left(\delta ^2+1\right) \left(s_1^2+s_2^2+s_3^2\right)+2
  X_B(2) s_2 \left(\delta ^2+1\right) \right. 
  \nonumber
  \\
  && \left.   \hspace{1cm}
  +\delta ^2 \left(X_B(3) \left(X_B(3)
   \left(s_1^2+s_2^2+s_3^2\right)+2 s_3\right)+2\right)+1\right]
\\
\beta_p&=&\frac{1}{\delta ^2}\left[X_B(2)^2 \left(\delta ^2+1\right) (s_1 (s_2+s_3)+s_2
   s_3)+X_B(2) \left(\delta ^2+1\right) (s_1+s_3)
\right. 
\nonumber
\\
&& \left. \hspace{1cm}     
   +X_B(3) \delta ^2
   (X_B(3) s_3 (s_1+s_2)+X_B(3) s_1
   s_2+s_1+s_2)\right]
\end{eqnarray} 
\begin{eqnarray}
\alpha_q&=&\frac{1}{2 \delta ^2}\left[W_B(1)^2 (2 \eta +1) \left(d_1^2+d_2^2+d_3^2\right) + 2  W_B(1)  d_1 (2 \eta +1)
\right. 
   \nonumber
   \\
   && \left.  \hspace{1cm}
- 2 W_B(1)\delta  \left(W_B(2)
   \left(d_1^2+d_2^2+d_3^2\right)+d_2\right) \right. 
   \nonumber
   \\
   && \left.  \hspace{1cm}
 + 2 W_B(1)\eta  d_1^2 \delta  (W_B(3)-2 W_B(2))
   \right. 
   \nonumber
   \\
   && \left.  \hspace{1cm}
      + 2 W_B(1)\eta\delta d_2^2
   (W_B(3)-2 W_B(2))\right. 
   \nonumber
   \\
   && \left.  \hspace{1cm}
   +  2 W_B(1)\eta\delta d_3^2 (W_B(3)-2 W_B(2)) 
   \right. 
   \nonumber
   \\
   && \left.  \hspace{1cm}
      + 2 W_B(1)\eta\delta (- 2 d_2+d_3) + (2\eta+1)  -d_2 \delta W_B(2) 
\right. 
   \nonumber
   \\
   && \left.  \hspace{1cm}
   +\delta 
   W_B(2)^2 (2 \eta +1) \delta  \left(d_1^2+d_2^2+d_3^2\right)-2\delta +d_1\delta W_B(2)\delta\right. 
   \nonumber
   \\
   && \left.  \hspace{1cm}
+\delta W_B(2) \eta  \left(\delta  \left(W_B(3)
   \left(d_1^2+d_2^2+d_3^2\right)-2 d_2+d_3\right)+2
   d_1\right) \right. 
   \nonumber
   \\
   && \left.  \hspace{1cm}
+\delta^2  W_B(3)^2 (2 \eta +1)
   \left(d_1^2+d_2^2+d_3^2\right)+2 \delta W_B(3) d_1 \eta
   \right. 
   \nonumber
   \\
   && \left.  \hspace{1cm}
   +2 W_B(3) \delta^2 (-d_2 \eta +2 d_3
   \eta +d_3)+4 \eta \delta^2+2\delta^2\right]
  \\
\beta_q&=&\frac{1}{\delta ^2} \left[W_B(1)^2 (2 \eta +1) (d_1 (d_2+d_3)+d_2 d_3)- W_B(1)d_1 (\eta +1) \delta 
\right.
   \nonumber
   \\
   && \left. \hspace{1cm} 
-2 W_B(1) W_B(2) d_2 (2 \eta +1) \delta  (d_1+d_3)
\right.
   \nonumber
   \\
   && \left. \hspace{1cm} 
+W_B(1) d_2 \eta  (\delta  (2
   W_B(3) (d_1+d_3)+1)+2)+W_B(1) d_2 
   \right.
   \nonumber
   \\
   && \left. \hspace{1cm} 
   -2 W_B(1) d_3 \eta  (2 W_B(2) d_1 \delta
   -W_B(3) d_1 \delta +\delta -1)\right.
   \nonumber
   \\
   && \left. \hspace{1cm} 
    -2 W_B(1) d_3 W_B(2) d_1 \delta - W_B(1) d_3(\delta   -1)
   \right.
   \nonumber
   \\
   && \left. \hspace{1cm} 
   +\delta W_B(2)^2 (2 \eta +1) \delta 
   (d_1 (d_2+d_3)+d_2 d_3)  \right.
   \nonumber
   \\
   && \left. \hspace{1cm} 
   -\eta W_B(2) \delta^2  d_1
   (2 W_B(3) (d_2+d_3)-1) 
   \right.
   \nonumber
   \\
   && \left. \hspace{1cm}    -\eta W_B(2) \delta^2 \left(2 W_B(3) d_2 d_3+d_2-2 d_3\right) \right.
   \nonumber
   \\
   && \left. \hspace{1cm} 
   -2 \eta W_B(2) \delta    (d_2+d_3)
   \right.
   \nonumber
   \\
   && \left. \hspace{1cm}    
   +d_1 W_B(2) \delta^2 -  W_B(2) \delta d_2+  W_B(2) \delta d_3 (\delta -1)  \right.
   \nonumber
   \\
   && \left. \hspace{1cm} 
   +W_B(3)^2 \delta  (2  \eta +1) \delta  (d_3 (d_1+d_2)+d_1 d_2)
     \right.
   \nonumber
   \\
   && \left. \hspace{1cm}
   +W_B(3)  \delta \eta 
   (\delta  (d_1+2 d_2-d_3)+d_2+d_3)
     \right.
   \nonumber
   \\
   && \left. \hspace{1cm}
   +W_B(3) \delta^2 
   (d_1+d_2)-\eta  \delta  (\delta +1)-\delta\right]
\end{eqnarray} 
\begin{eqnarray}
\gamma_{A^\prime}&=& \frac{C^2}{2}
\ ,
\\
\mu_{A^\prime} &=& \frac{1}{2 \text{C}^2 \delta ^2}\left[\delta ^2+2 \eta  ((\delta -1) \delta +1)+1\right]
\ , 
\end{eqnarray}
\begin{eqnarray}
\gamma_{B^\prime}&=& \frac{s_B^2}{2
   \delta ^2} \left[ \delta ^2 \left(X_B(2)^2+X_B(3)^2\right)+X_B(2)^2\right]
\ ,
\\
\omega_{B^\prime}&=& \frac{d_B^2}{2 \delta ^2} \left[ W_B(1)^2 (2 \eta +1)-2 W_B(1) \delta  (2 W_B(2) \eta
   +W_B(2)-W_B(3) \eta )
   \right.
   \nonumber
   \\
   && \left. \hspace{1cm} 
   +2 \delta ^2  \eta  \left(W_B(2)^2-W_B(2)
   W_B(3)+W_B(3)^2\right)\right.
   \nonumber
   \\
   && \left. \hspace{1cm} + \delta^2 \left( W_B(2)^2+W_B(3)^2 \right) \right]
\end{eqnarray} 
In particular if we study the behavior of these quantities around the limit $\delta\rightarrow0$ we find:
\begin{eqnarray}
g&\sim & \ \mathcal{O}(\delta^{-1}) \ , 
\\
X_A(1) &\sim & \ \mathcal{O}(\delta^{-1/2}) \ , 
\\
W_A(1)&\sim & \ \mathcal{O}(\delta^{1/2}) \ , 
\\
W_A(2)&\sim & \ \mathcal{O}(\delta^{-1/2}) \ , 
\\
X_B(1) &\sim & \ \mathcal{O}(\delta^{-1/2}) \ , 
\\
X_B(2)&\sim & \ \mathcal{O}(\delta^{1/2}) \ , 
\\
X_B(3)&\sim & \ \mathcal{O}(\delta^{+1/2}) \ , 
\\
W_B(1) &\sim & \ \mathcal{O}(\delta^{+1/2}) \ , 
\\
W_B(2)&\sim & \ \mathcal{O}(\delta^{-1/2}) \ , 
\\
W_B(3)&\sim & \ \mathcal{O}(\delta^{+1/2})\ , 
\\
\Omega&\sim & \ \mathcal{O}(\delta^{0}) \ , 
\end{eqnarray}
which implies:
\begin{eqnarray}
\alpha_p &\sim & \ \mathcal{O}(\delta^{-2}) \ , 
\\
\beta_p &\sim & \ \mathcal{O}(\delta^{-1}) \ , 
\\
\alpha_q &\sim & \ \mathcal{O}(\delta^{-2}) \ , 
\\
\beta_q &\sim & \ \mathcal{O}(\delta^{-1}) \ , 
\\
\gamma_{A^\prime} &\sim & \ \mathcal{O}(\delta^{-1}) \ , 
\\
\mu_{A^\prime} &\sim & \ \mathcal{O}(\delta^{-1}) \ , 
\\ 
\gamma_{B^\prime} &\sim & \ \mathcal{O}(\delta^{-1}) \ , 
\\
\mu_{B^\prime} &\sim & \ \mathcal{O}(\delta^{-1}) \ .
\end{eqnarray}
The Taylor series coefficients of the dominant contribution in the limit $\delta\rightarrow0$ of the previous equations can be found plotted in figures  \eqref{divergences} and  \eqref{optimizes}.


\begin{thebibliography}{99}                                                                                               %


\bibitem {EE1}L. Bombelli, R. K. Koul, J. Lee, and R. Sorkin, Phys. Rev. D 34,
373 (1986).

\bibitem {EE2}M. Srednicki, Phys. Rev. Lett. \textbf{71}, 666 (1993).

\bibitem {EE3}J. Eisert, M. Cramer, and M. B. Plenio, Rev. Mod. Phys.
\textbf{82}, 277 (2010).

\bibitem {EH1}A. Valentini, Physics Letters A \textbf{153}, 321 (1991).

\bibitem {EH2}B. Reznik, Foundations of Physics \textbf{3}3, 167 (2003).

\bibitem {EH3}E. Martin-Martinez, E. G. Brown, W. Donnelly, and Achim Kempf,
Phys. Rev. A \textbf{88}, 052310 (2013).

\bibitem {EH4}G. Salton, R. B. Mann, and N. C. Menicucci, New J. Phys.
\textbf{17}, 035001 (2015).

\bibitem {QO}M. O. Scully and M. S. Zubairy, \textit{Quantum Optics,
}Cambridge University Press (1997).

\bibitem {SC}A. Wallraff, D. I. Schuster, A. Blais, L. Frunzio, R. S. Huang,
J. Majer, S. Kumar, S. M. Girvin, and R. J. Schoelkopf, Nature \textbf{431},
162 (2004).

\bibitem {QHS}D. Yoshioka, \textquotedblleft The Quantum Hall
Effect\textquotedblright, Springer (2002).

\bibitem {HSU}M. Hotta, R. Sch\"{u}tzhold, and W. G. Unruh, Phys. Rev. D
\textbf{91}, 124060.

\bibitem {h}S. W. Hawking, Phys. Rev. D \textbf{14}, 2460, (1976).

\bibitem {TYH}J. Trevison, K. Yamaguchi, and M. Hotta, ``General Entangled Partner in Quantum Field Theory'', arXiv:1807.03467 

\bibitem {cc1}D. Gross and J. Eisert, Phys. Rev. Lett. \textbf{98}, 220503 (2007).

\bibitem {cc2}J. M. Cai, W. D\"{u}r, M. Van den Nest, A. Miyake, and H. J.
Briegel, Phys. Rev. Lett. \textbf{103}, 050503 (2009).

\end{thebibliography}
\end{document}